\documentstyle[11pt,IAU207_pasp,twoside,psfig]{article}

\begin{document}

\title{Formation of Twin Clusters in a Galactic Tidal Field}
\author{Christian Theis}
\affil{Institut f.\ Theoretische Physik und Astrophysik, Universit\"at Kiel,
24098 Kiel, Germany, email: theis@astrophysik.uni-kiel.de}


\begin{abstract}
     The formation of globular clusters is still an unsolved problem. Though
  most scenarios assume a massive molecular cloud as the progenitor, 
  it is unclear, how the cloud is transformed into a star cluster.
  Here a scheme of supernova (SN) induced cluster formation is investigated. 
  In this scenario the expanding SN shell accumulates the mass of the cloud.
  This is accompanied by fragmentation resulting in
  star formation in the shell. If this stellar shell expands sufficiently slow,
  its self-gravity leads to a recollapsing shell, by this forming 
  one or several stellar clusters.

    In this paper N-body simulations of collapsing shells 
  moving on circular orbits in a galactic potential
  are presented. It is shown that typical shells ($10^5$ M$_\odot$, 30 pc)
  evolve to twin clusters in the galactocentric distance range between 
  3 and 11 kpc.
  Their masses show a strong radial trend: on orbits 
  inside 5 kpc both clusters have almost equal mass. Outside 5 kpc the more 
  massive twin cluster contains about 55\% of the shell's mass, whereas the 
  mass of the smaller decreases linearily to 15\% at 11 kpc. Outside 11 kpc the 
  collapsing shells end up in a single cluster. Inside 3 kpc the shells are
  tidally disrupted and only fragments substantially less massive than the 
  initial shell survive.
\end{abstract}

\vspace*{-1cm}

\section{Introduction}

  Several scenarios for the formation of globular cluster 
have been suggested, e.g.\ the collapse of giant
molecular clouds (GMC) or the collision of molecular clouds 
(e.g.\ Fall \& Rees 1985, Murray \& Lin 1990, Fujimoto \& Kumai 1997). 
A common characteristic of 
all these scenarios is, that the clusters are formed from smooth
gaseous distributions which are transformed into stars.
However, this assumption requires short formation timescales and unusually 
high star formation efficiencies in order to end up with a gravitationally 
bound cluster.
  An alternative model suggested by Brown et al.\ (1991) can overcome these
difficulities: their scenario
starts with an OB-association exploding near the center 
of a molecular cloud. The expanding shell sweeps up the cloud material and in 
a later stage the expansion can be decelerated or stopped by both,
the accumulated mass and the external 
pressure of the ambient hot gas. The shell itself undergoes fragmentation
and, finally, forms stars. If these stars are gravitationally bound, they 
will recollapse, by this forming a globular cluster.
 
  Though a discrimination between the scenarios by means of hydrodynamical 
simulations (starting from first principles) is far out of reach at the
moment, one can study different evolutionary stages.
E.g.\ Theis (2000) compared in a series
of N-body simulations the collapse of thin stellar shells
and homogeneous spheres in a galactic tidal field. These calculations 
were performed for circular and
eccentric orbits, but with a constant apogalacticon of 5 kpc. It was found that
collapsing shells preferably end in multiple systems, mainly twins, whereas
homogeneous spheres either form single clusters or become completely disrupted.
The twins might survive for -- at least -- several galactic revolutions, and
some of them show kinematical peculiarities like counter-rotating cores.

   In this paper I will address the question, how the formation of twins
depends on the galactocentric distance, i.e.\ the strength of the galactic 
tidal field. 

\section{Numerical Models}

   The simulations start with a thin, spherical shell of a
mass M$_{\rm tot} = 10^5$ M$_\odot$, an outer radius
of 30 pc and a thickness of 3 pc. The shell is initially at rest in the sense
that there is no overall expansion or contraction of the shell with respect
to its center. 
The potential of the Galaxy is modelled
by an isothermal halo with a circular speed of 220 km\,s$^{-1}$. In all 
simulations the initial velocity of the shell corresponds to a circular
orbit. The calculations are performed with $N=10^4$ particles 
using a GRAPE3 board. 

\begin{figure}[t]
  \centerline{\hbox{
  \psfig{figure=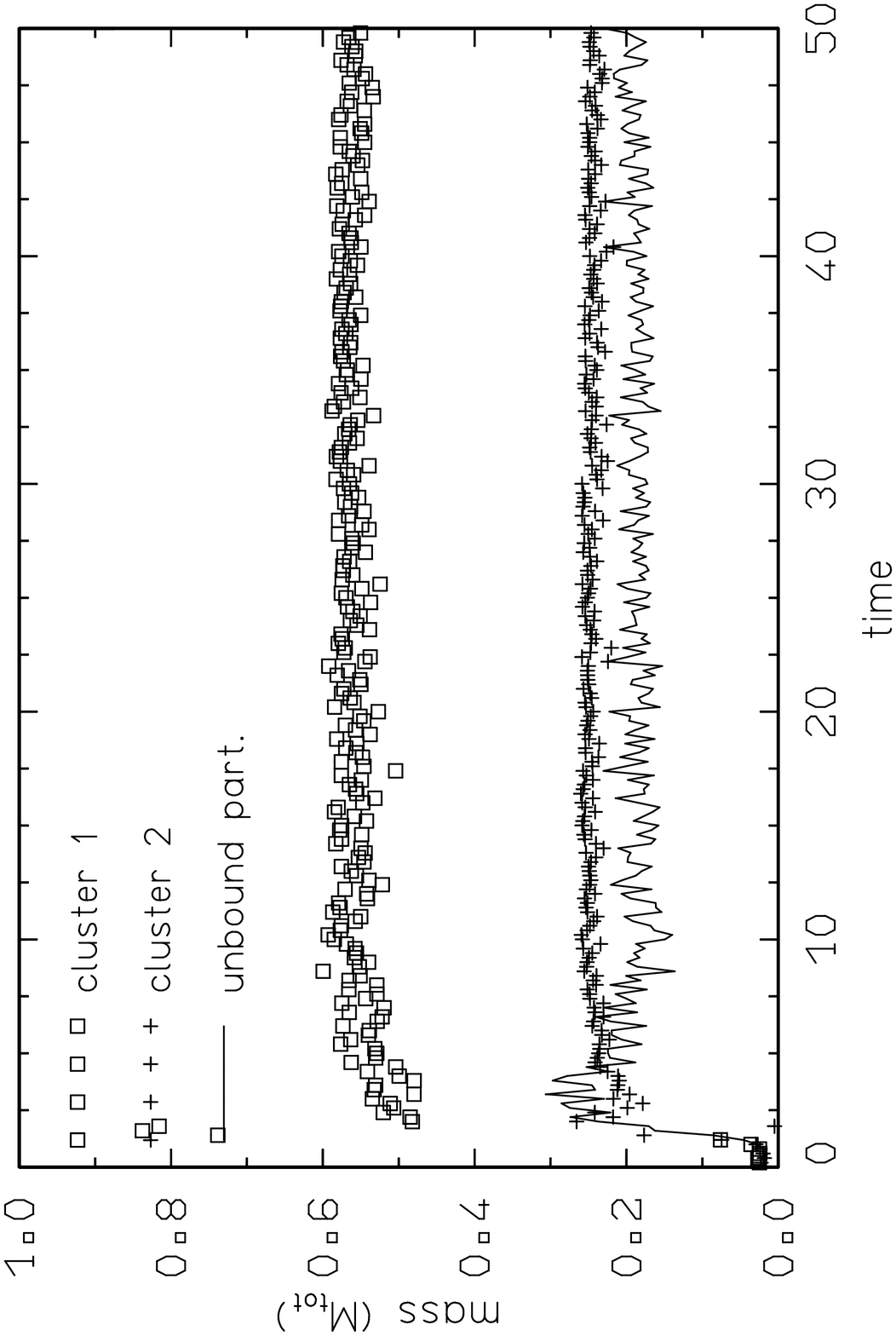,width=6.5cm,angle=270}
  \psfig{figure=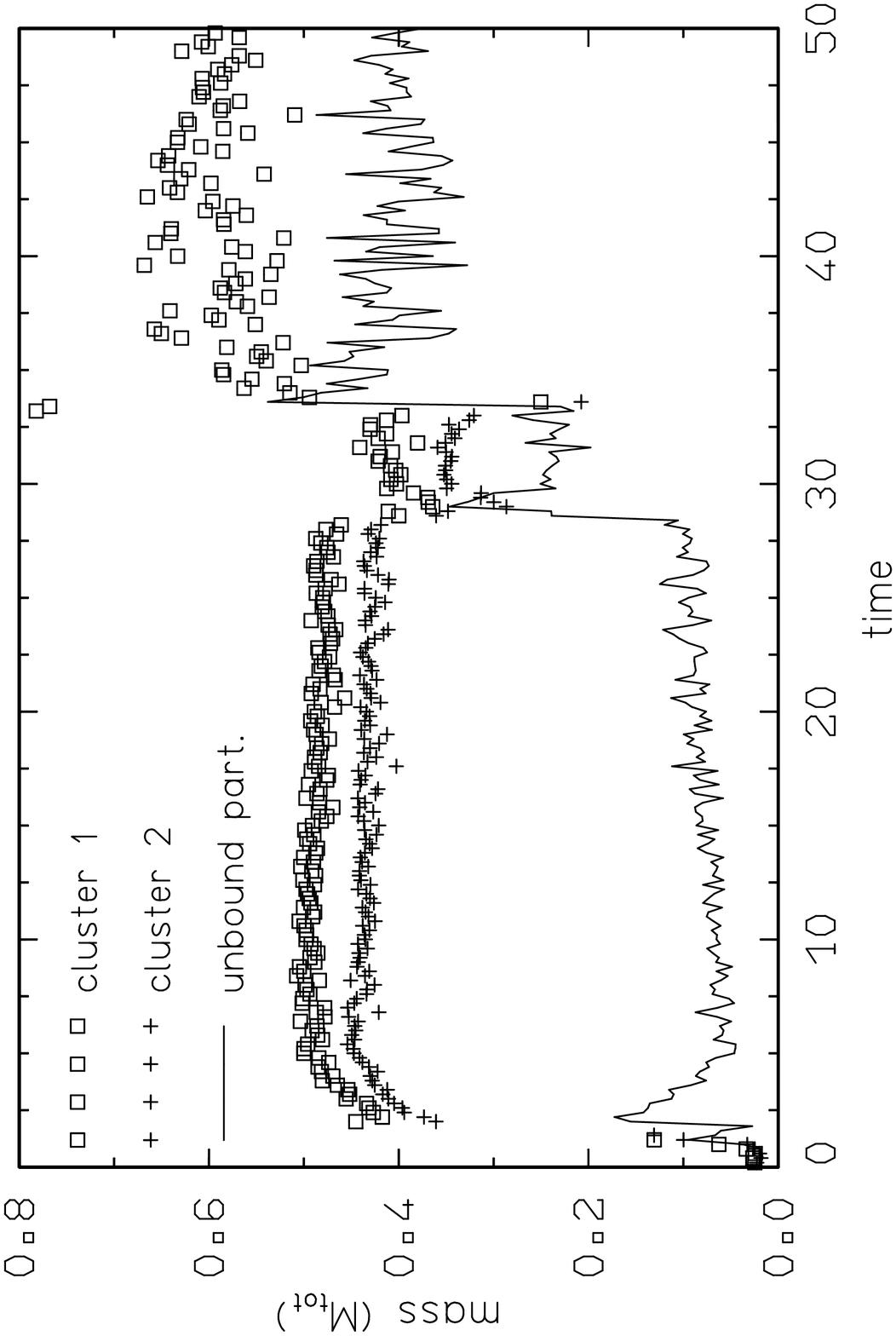,width=6.5cm,angle=270}
  }}
  \caption{Temporal evolution of the masses of the twin clusters (signs) and 
     the mass of unbound particles (solid line). Shown are results for 
     circular orbits at 
     a galactocentric distance of 10 kpc (left) and 5 kpc (right). The twins
     in the right panel collide at $t \approx 28.6$ and merge at 
     $t \approx 33.4$. The total mass in each simulation is
     10$^5$ M$_\odot$. The time unit is 7.7 Myr.
   }
\end{figure}

   Fig.\ 1 shows the temporal evolution of the masses of the two largest
clusters for galacatocentric radii of 5 and 10
kpc. At 10 kpc two clusters with a mass ratio of 5:2 are formed after 20 Myr.
Both clusters survive until the end of the simulation ($\sim 400$ Myr).
20\% of the stars initially residing in the shell became unbound. A 
slightly different behaviour is seen for the model starting at 5 kpc:
Again twins are formed, but they have almost equal mass. After 210 Myr 
they collide and, finally, they merge. Both events are clearly reflected in
step-like increases of the number of unbound particles which amounts
finally to 40\% of the mass of the shell (Fig.\ 1, right).

\begin{figure}
  \centerline{\vbox{
  \psfig{figure=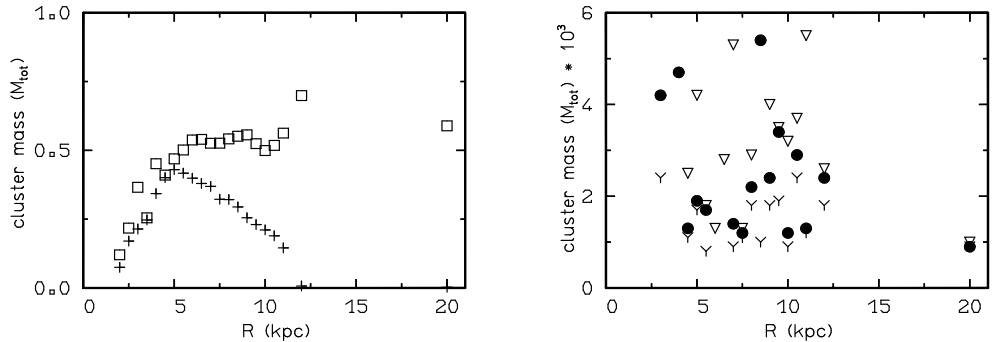,width=13cm,angle=270}
  }}
  \caption{Cluster mass vs.\ galactocentric distance. Shown are the masses
    of the five most massive clusters of each simulation 18.6 Myr after the 
    collapse. The distribution is unaffected by later merging of clusters. 
    The clusters
    are sorted by mass (left: most massive clusters (box, plus), right: smaller
    clusters (triangle, filled circle, Y)).
   }
\end{figure}

  The masses of the formed clusters exhibit a clear radial trend (Fig.\ 2):
The mass of the largest cluster increases almost linearily from 2 to
5.5 kpc reaching a plateau of about 55\% of the total mass. Beyond 11 kpc
only one massive cluster is formed. About 40\% of the stars are
not bound in clusters. Inside 3 kpc, a large set of small clusters
is formed instead of a dominating pair of clusters. The mass ratio $q$
between the two most massive clusters also shows two regimes: Below
5 kpc the masses are almost identical, whereas outside 5 kpc $q$ increases
up to $\sim 4:1$ at 11 kpc.
  In addition to the large clusters ($>10^4$ M$_\odot$), typically several
small ($<10^3$ M$_\odot$), gravitationally bound clusters are formed.
Contrary to the large clusters, their mass shows no trends with galactocentric 
distance.

  The simulations demonstrate that twin formation is expected over a large
radial range. About 1/3 of these twins merge within 400 Myr after their
formation. The surviving twins are characterized by large
spatial separations which makes them less likely to undergo a subsequent
merger. Therefore, twin globulars might still exist in the Milky Way, but they
can be unidentified as twins due to their large separation. Since they should
share common orbits and metallicities, they could be found by proper
determinations of globular cluster orbits.

\acknowledgments
The author is grateful to the {\it Deutsche Forschungsgemeinschaft} (DFG)
for the travel support under grant TH511/3--1. The analysis of the
cluster sizes has been performed with the SKID program kindly made available
by the NASA HPCC ESS group at the University of Washington.

\end{document}